\documentclass[11pt]{article}
\usepackage{booktabs}
\usepackage{multirow}
\usepackage{xcolor}
\usepackage[most]{tcolorbox}
\usepackage[final]{acl}
\usepackage{siunitx}
\usepackage{times}
\usepackage{latexsym}
\usepackage{enumitem}
\usepackage[T1]{fontenc}
\usepackage{amsfonts}
\usepackage{amsmath}

\usepackage[utf8]{inputenc}
\usepackage{microtype}
\definecolor{darkgreen}{RGB}{0, 100, 0}
\definecolor{MyTitleBg}{RGB}{220, 215, 235}  
\definecolor{MyContentBg}{RGB}{250, 248, 255} 
\definecolor{MyFrame}{RGB}{220, 215, 235}     
\newtcolorbox{PromptBox}[2][]{
    enhanced jigsaw,                   
    breakable,
    colback=MyContentBg,         
    colframe=MyFrame,            
    colbacktitle=MyTitleBg,      
    coltitle=black,              
    fonttitle=\bfseries\small,   
    title={#2},                  
    arc=3mm,                     
    boxrule=0.8pt,               
    titlerule=0pt,               
    boxsep=1mm,
    left=2mm, right=2mm, top=2mm, bottom=2mm,
    toptitle=1.5mm, bottomtitle=1.5mm, 
    width=\columnwidth,          
    #1                           
}

\definecolor{DarkTitleBg}{RGB}{50, 50, 55}      
\definecolor{LightContentBg}{RGB}{245, 245, 248} 
\definecolor{DarkFrame}{RGB}{50, 50, 55}        

\newtcolorbox{BlackPromptBox}[2][]{
    enhanced jigsaw, breakable,      
    colback=LightContentBg,          
    colframe=DarkFrame,              
    colbacktitle=DarkTitleBg,        
    coltitle=white,                  
    fonttitle=\bfseries\small,
    title={#2},
    arc=3mm,
    boxrule=0.8pt,
    titlerule=0pt,                   
    boxsep=1mm,
    left=2mm, right=2mm, top=2mm, bottom=2mm,
    toptitle=1.5mm, bottomtitle=1.5mm,
    width=\columnwidth,              
    #1
}

\usepackage{inconsolata}

\usepackage{graphicx}

\newcommand{\wrt}{\emph{w.r.t. }}
\title{Scattered Hypothesis Generation for Open-Ended Event Forecasting
}

\author{
 \textbf{He Chang\textsuperscript{$\spadesuit$}},
 \textbf{Zhulin Tao \textsuperscript{$\spadesuit$}\footnotemark[1]},
 \textbf{Lifang Yang\textsuperscript{$\spadesuit$}},
 \textbf{Xianglin Huang\textsuperscript{$\spadesuit$}},
 \textbf{Yunshan Ma\textsuperscript{$\diamondsuit$}},
\\
 \textsuperscript{$\spadesuit$}Communication University of China,
 \textsuperscript{$\diamondsuit$}Singapore Management University,
\\
 \texttt{\{hechangcuc,yanglifang,huangxl\}@cuc.edu.cn} \\
 \texttt{taozhulin@gmail.com, ysma@smu.edu.sg}
}
\begin{document}
\maketitle
\begin{abstract}
Despite the importance of open-ended event forecasting for risk management, current LLM-based methods predominantly target only the most probable outcomes, neglecting the intrinsic uncertainty of real-world events. To bridge this gap, we advance open-ended event forecasting from pinpoint forecasting to \textit{scatter forecasting} by introducing the proxy task of hypothesis generation. This paradigm aims to generate an inclusive and diverse set of hypotheses that broadly cover the space of plausible future events. To this end, we propose SCATTER, a reinforcement learning framework that jointly optimizes inclusiveness and diversity of the hypothesis. Specifically, we design a novel hybrid reward that consists of three components: 1) a \textit{validity reward} that measures semantic alignment with observed events, 2) an \textit{intra-group diversity reward} to encourage variation within sampled responses, and 3) an \textit{inter-group diversity reward} to promote exploration across distinct modes. By integrating the validity-gated score into the overall objective, we confine the exploration of wildly diversified outcomes to contextually plausible futures, preventing the mode collapse issue. Experiments on two real-world benchmark datasets, i.e., OpenForecast and OpenEP, demonstrate that SCATTER significantly outperforms strong baselines. Our code is available at \url{https://github.com/Sambac1/SCATTER}.
\end{abstract}
\footnotetext[1]{Corresponding author}
\section{Introduction}

Event forecasting plays a critical role in risk management, public policy, and strategic decision-making by enabling proactive responses to future socio-economic and geopolitical developments rather than reactive measures~\cite{lee2025advancing, superforecastor, forecastNN}. With recent advances in large language models (LLMs)~\cite{deepseekr1, gpt4} and post-training techniques, LLM-based event forecasting has emerged as a promising approach. Existing research~\cite{SurveyAdvances, TEFevalution, ai-augmented} typically falls into two primary paradigms: discriminative forecasting and open-ended forecasting. The former~\cite{BTF, forecastbench, Prophet} typically formulates forecasting as a closed-set decision problem such as multiple-choice questions or binary predictions. In contrast, open-ended forecasting~\cite{openforecast, futureX} allows LLMs to generate free-form natural language predictions conditioned on historical context. Owing to its capacity to model complex event dynamics and provide detailed generative insights, open-ended event forecasting has recently garnered substantial attention.
\begin{figure}[t]
  \includegraphics[width=\columnwidth]{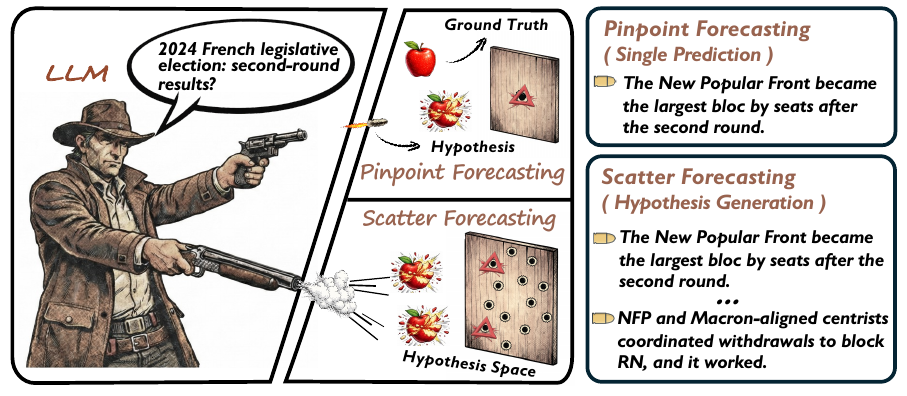}
  \caption{The analogy between shooting and event forecasting: pinpoint forecasting (single prediction) versus scatter forecasting (hypothesis generation).}
  \vspace{-0.15in}
  \label{fig:intro}
\end{figure}

However, existing open-ended event forecasting methods~\cite{openep, TCELong, forestcast} largely adhere to a single prediction paradigm. As shown in the analogy of Figure~\ref{fig:intro}, this paradigm resembles pinpoint shooting with a single bullet, where the forecasting process yields only one most probable outcome.
However, in the real world, the possible future developments of an ongoing event often involve multiple plausible events rather than a single outcome. 
Consequently, pinpoint forecasting, which restricts outputs to a narrow subset of plausible futures, fails to capture the intrinsic uncertainty of real-world events~\footnote{Further evidence can be seen in Section~\ref{model_study}.}.
To bridge this gap, we introduce \textit{scatter forecasting}. Analogous to scatter shooting, where each shot covers an area (i.e., a set of points), scatter forecasting produces a set of outcomes, each representing a hypothesis about the future event.
Accordingly, open-ended event forecasting can be reformulated as the proxy task of \textit{hypothesis generation}: given a certain context, an LLM-based forecasting model generates a set of distinct hypotheses that maintain both \textit{inclusiveness} and \textit{diversity}, thereby covering the space of plausible future events.

Although hypothesis generation is promising, simultaneously ensuring both inclusiveness and diversity of generated hypotheses remains challenging.
A straightforward approach~\cite{wisdom, app, openep} is to prompt LLMs to generate multiple non-redundant hypotheses; however, most existing LLMs, primarily trained via supervised fine-tuning, struggle to consistently produce distinct yet valid long-tail events~\cite{understanding,LLMlongtail}.
Alternatively, reinforcement learning (RL)~\cite{RLintro, RLHF, ppo}, such as Group Relative Policy Optimization (GRPO)~\cite{deepseekmath, dapo}, appear to be a promising solution.
Nevertheless, optimizing for both inclusiveness and diversity is inherently conflicting. On the one hand, emphasizing inclusiveness tends to favor conservative hypotheses and is prone to mode collapse~\cite{modecollapse}, ultimately reducing the output to a single outcome. On the other hand, emphasizing diversity, in contrast, often yields noisy gradients and unstable optimization, thereby hampering convergence~\cite{diversity-aware}. 
This tension is further exacerbated by the irreversibility of real-world events and the scarcity of counterfactual data, which hinder the acquisition of high-quality training data~\cite{aicollapse}.

To address these challenges, we propose \textbf{SCATTER}: a reinforcement learning framework for scattered hypothesis generation. Based on GRPO, SCATTER jointly optimizes inclusiveness and diversity with a hybrid reward that confines exploration to contextually plausible futures. Given the absence of objective gold standards and the sparsity of observable real-world events, we first introduce a validity reward based on embedding similarity. This serves as a coarse-grained yet effective semantic proxy to align generated hypotheses with available factual evidence. Conditioned on validity-gated score, we design two diversity rewards: an intra-group reward to encourage variation within individual samples and an inter-group reward to promote exploration across distinct plausible modes and mitigate mode collapse. By down-weighting diversity rewards for non-factual hypotheses, our approach prevents the reward hacking triggered by trivial novelty, thereby anchoring the learning process toward a diverse yet plausible hypothesis space. We conduct extensive experiments on two real-world benchmark datasets OpenForecast~\cite{openforecast} and OpenEP~\cite{openep}, and the results demonstrate that our method significantly outperforms strong baselines.
Our contributions are summarized as follows: 
\begin{itemize}[leftmargin=*]
\item We formulate open-ended event forecasting as a \textit{hypothesis generation} problem, advancing from single prediction to scatter forecasting. 
\item We propose \textbf{SCATTER}, a reinforcement learning framework that jointly optimizes inclusiveness and diversity via a hybrid reward design.
\item Empirical results demonstrate that SCATTER significantly outperforms strong baseline methods. \end{itemize}
\section{Related Work}

\subsection{LLM-Based Event Forecasting}
Event forecasting~\cite{promise,SCTc-TE,context} aims to predict future outcomes from historical context and has traditionally been approached with structured statistical and neural methods. More recently, LLMs have been integrated directly into forecasting pipelines~\cite{TGL-LLM, mm-forecast,LLMforecastsurvery}. Broadly, prior work~\cite{app,forecastqa, thinktank} falls into two paradigms: discriminative forecasting and generative forecasting.
Discriminative approaches~\cite{TEFevalution,forecastbench} cast forecasting as closed-set prediction (e.g., binary or multiple-choice). While competitive on constrained question types, these methods inherently restrict forecasts to a predefined outcome space.
Generative approaches~\cite{openforecast,TCELong}, in contrast, treat forecasting as open-ended prediction, allowing LLMs to produce free-form descriptions of future events conditioned on context. However, most existing methods~\cite{futureX,openep} primarily emphasize dataset construction and baseline evaluation pipelines, with modeling objectives still largely restricted to generating a single prediction per query. In practice, real-world dynamics are complex and multi-modal, and a single deterministic prediction cannot capture the range of plausible future trajectories. These considerations motivate reframing open-ended event forecasting from single prediction to hypothesis generation.

\subsection{Post-Training for LLMs}
Post-training methods~\cite{peft, sftrl}, including supervised fine-tuning (SFT), reinforcement learning~\cite{DPO} and preference-based optimization~\cite{instruct_training}, are widely used to steer pretrained LLMs toward desired downstream behaviors.
The rapid advancement of reinforcement learning~\cite{deepseeknature}, especially GRPO, has garnered significant attention, particularly in domains such as mathematics and coding. However, while RL enhances sampling efficiency towards correct paths, current training paradigms rarely elicit fundamentally new reasoning patterns, rendering models prone to mode collapse~\cite{modecollapse}. This tendency is particularly detrimental in open-ended forecasting, where practitioners require a diverse set of credible scenarios rather than a single canonical prediction. Although recent research~\cite{seedgrpo, evolving} has begun to investigate exploration within RL, it remains confined to verifiable domains. These approaches are insufficient for open-ended forecasting because they rely on deterministic verification signals (e.g., compilers or solvers) to prune search spaces, signals that are inherently absent in forecasting tasks where ambiguity prevails and \textit{correctness} is probabilistic rather than binary.


\section{Preliminary}
\label{sec:prelim}
\subsection{Problem Formulation}
We reformulate open-ended event forecasting as a hypothesis generation task to better accommodate the intrinsic uncertainty of the real world. Formally, let $\mathcal{D} = \{(\mathcal{C}_i, \mathcal{Q}_i, \mathcal{G}_i)\}_{i=1}^N$ denote a dataset of $N$ samples. For each sample, the input context $x = (\mathcal{C}, \mathcal{Q})$ comprises the historical background $\mathcal{C}$ and the specific forecasting question $\mathcal{Q}$, while $\mathcal{G}$ represents the ground-truth outcome. See Appendix~\ref{app:case_study} for a detailed example. We aim to train an LLM policy $\pi_\theta$ to serve as the event forecaster. Given a query context $x$, the policy samples a set of responses $\{y_k\}_{k=1}^K$, where each response $y_k=\{h_{k,1},\dots,h_{k,M}\}$ contains $M$ hypotheses describing plausible future events.

\subsection{Optimization with GRPO} We optimize the policy $\pi_\theta$ using GRPO. Unlike standard PPO, which requires a parametric value function, GRPO utilizes the mean reward of a sampled group as the baseline, reducing computational overhead. Specifically, for each context $x$, we sample a group of $G$ responses $\{y_1, \dots, y_G\}$ from the old policy $\pi_{\theta_{\text{old}}}$.
The optimization objective is defined as follows:
\begin{equation}
\label{eq:grpo_full}
\begin{aligned}
    \mathcal{J}_{\text{GRPO}}(\theta) = \mathbb{E}_{x, \varepsilon} \Bigg[ & \frac{1}{G} \sum_{i=1}^{G} \min \Big( \rho_i \hat{A}_i, \\
    & \quad \operatorname{clip}(\rho_i, 1-\epsilon, 1+\epsilon) \hat{A}_i \Big) \\
    & - \beta \mathbb{D}_{\text{KL}}(\pi_\theta || \pi_{\text{ref}}) \Bigg],
\end{aligned}
\end{equation}
where $\rho_i(\theta) = \frac{\pi_\theta(y_i|x)}{\pi_{\theta_{\text{old}}}(y_i|x)}$ is the importance sampling ratio. The advantage $\hat{A}_i$ is computed by normalizing the rewards within the group:
\begin{equation}
\hat{A}_i = \frac{r(y_i, x) - \mu_r}{\sigma_r + \delta},
\end{equation}
where $r(y_i, x)$ is the reward for the response $y_i$, and $\mu_r, \sigma_r$ denote the mean and standard deviation of the rewards within the sampled group, respectively. $\delta$ is a small constant for numerical stability.
\section{Our Approach: SCATTER}

\begin{figure*}[t]
  \centering
  \includegraphics[width=\textwidth]{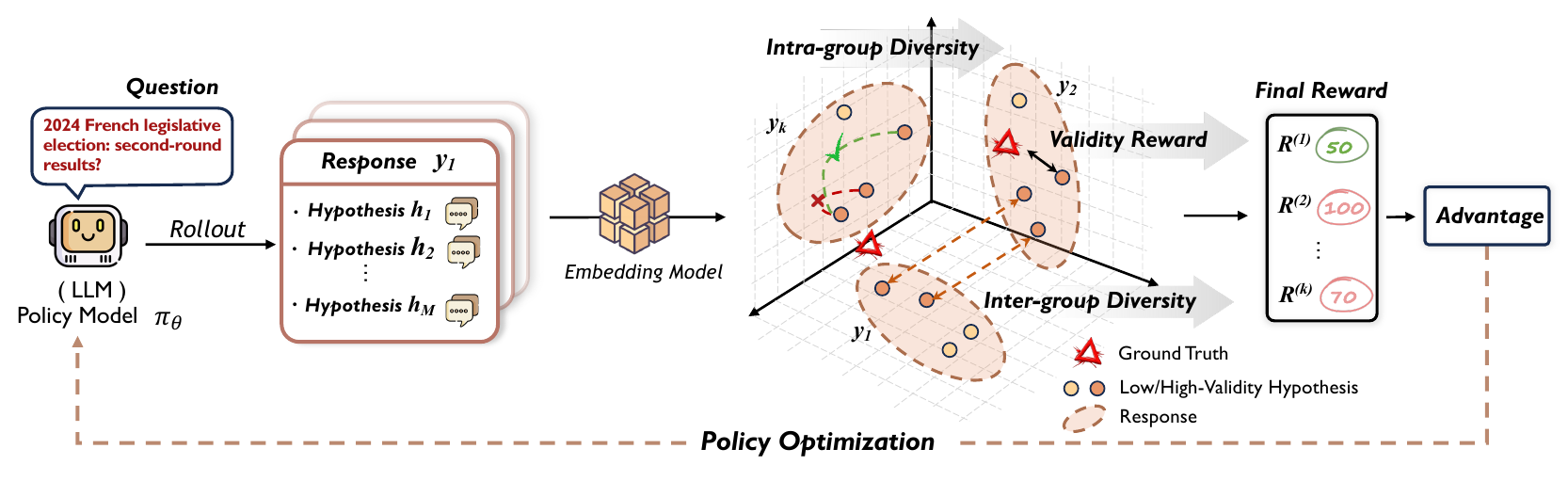}
\caption{Overall framework of SCATTER. A policy model (LLM) samples multiple hypothesis sets (rollouts) for a given query. These hypotheses are mapped into an embedding space to evaluate \textit{validity} (alignment with ground truth), \textit{intra-group diversity} within each response and \textit{inter-group diversity} across different responses. The resulting rewards are used for policy optimization, guiding the model to produce inclusive and diverse hypothesis sets.}
  \label{fig:framework}
\end{figure*}

We present our proposed approach SCATTER, which is a hybrid reward mechanism that aims to maximize diversity while maintaining inclusiveness, illustrated Figure~\ref{fig:framework}. Specifically, SCATTER consists of a validity reward and a validity-gated diversity mechanism. This architecture employs the validity signal to ensure factual alignment, while the gate regulates intra-group and inter-group diversity rewards to prevent reward hacking.
\subsection{Validity Reward}
\label{sec:validity_reward} 
To quantify the semantic correctness and alignment of the generated hypotheses, we define a validity score $q_{k,i}$ relative to ground truth set $\mathcal{G}_x$. Unlike deterministic tasks (e.g., math or code), open-ended event forecasting is not directly verifiable due to free-form ground truth. Therefore, we use a pretrained text embedding model to encode each hypothesis $h_{k,i}$ into a dense vector representation. Let $\mathbf{h}_{k,i}$ denote the embedding of $i$-th hypothesis in the $k$-th sampled response. We compute $q_{k,i}$ as the maximum cosine similarity between $h_{k,i}$ and the embeddings of ground truth events $\mathbf{g}$.
\begin{equation}
q_{k,m} =  \max_{\mathbf{g} \in \mathcal{G}x} \cos(\mathbf{h}_{k,i}, \mathbf{g}) .
\end{equation} 
This maximization implies a recall-oriented evaluation strategy: a hypothesis is considered valid if it aligns with at least one plausible future, mirroring the one-to-many nature of open-ended forecasting.

We then aggregate hypothesis-level validity scores to obtain a response-level validity reward by averaging over the $M$ hypotheses:
\begin{equation}
R_{\text{validity}}^{(k)} = \frac{1}{M} \sum_{i=1}^{M} q_{k,i}.
\end{equation}
While $R_{\text{validity}}^{(k)}$ promotes semantic correctness, optimizing this objective in isolation often causes the policy $\pi_\theta$ to converge to a single mode, resulting in mode collapse.





\subsection{Intra-Group Diversity Reward}

To incentivize diversity among generated hypotheses while preserving validity, we introduce a intra-group diversity reward. Specifically, we calculate the pairwise cosine dissimilarity between hypotheses, weighted by their respective validity scores:
\begin{equation}
\label{eq:intra}
\begin{aligned}
    R_{\text{intra}}^{(k)} = \frac{2}{M(M-1)} \sum_{i < j} & \Big[ \bigl(1 - \cos(\mathbf{h}_{k,i}, \mathbf{h}_{k,j})\bigr) \\
    & \sqrt{ q_{k,i} \, q_{k,j} } \Big].
\end{aligned}
\end{equation}
Here, the term $1 - \cos(\cdot, \cdot)$ encourages geometric separation in the embedding space, while $\sqrt{ q_{k,i} q_{k,j} }$ serves as a soft constraint. Crucially, this constraint penalizes pairs involving low-validity hypotheses, preventing the model from inflating diversity with semantically irrelevant outliers. As a result, optimization favors a hypothesis set that is both inclusive and diverse.

\subsection{Inter-Group Diversity Reward}

Beyond local diversity, global exploration is essential to prevent the policy from collapsing into a single mode across different sampling groups. We define an inter-group diversity reward to quantify the unique contribution of each response relative to the global hypothesis pool.
\paragraph{Importance Reweighting.} We first construct a probability distribution over hypotheses within response $k$ to prioritize high-validity hypotheses. The normalized importance weight $\omega_{k,i}$ is defined as:
\begin{equation}
\omega_{k,i} = \frac{q_{k,i}}{\sum_{j=1}^{M} q_{k,j} + \epsilon}.
\end{equation}
This ensures that the divergence metrics focus on the separation of plausible futures rather than distinguishing between noise.
\paragraph{Weighted Asymmetric Chamfer Distance.} To measure the directional diversity from response $k$ to response $l$, we employ a weighted variant of the Asymmetric Chamfer Distance~\cite{chamfer, chamferdensity}. Specifically, we calculate the expected minimum distance from the salient hypotheses in $k$ to the nearest neighbor in response $l$:
\begin{equation}
D(k \to l) = \sum_{i=1}^{M} \omega_{k,i} \left( 1 - \max_{j} \cos(\mathbf{h}_{k,i}, \mathbf{h}_{l,j}) \right).
\end{equation}
Intuitively, $D(k \to l)$ represents the coverage gap: it is maximized when the high-validity hypotheses in $k$ have no semantic counterparts in $l$.
\paragraph{Validity-Gated Diversity Reward}
We aggregate the diversity of response $k$ against all other responses $l \neq k$ via a leave-one-out average:
\begin{equation}
S_{\text{raw}}^{(k)} = \frac{1}{G-1} \sum_{l \neq k} D(k \to l).
\end{equation}
However, diversity alone is insufficient, as irrelevant hypotheses can inflate diversity without improving coverage. We therefore apply a validity-gated modulation, scaling the raw diversity reward by the response-level validity:
\begin{equation}
R_{\text{inter}}^{(k)} = S_{\text{raw}}^{(k)} \sqrt{\frac{1}{M} \sum_{i=1}^{M} q_{k,i}}.
\end{equation}
This gating mechanism ensures that the model is rewarded only for discovering novel and high-validity regions of the semantic space.

\subsection{Final Reward}
\label{sec:final_reward}

We aggregate the three components into a single composite scalar reward: 
\begin{equation} 
\label{eq:finalreward} R^{(k)} = R_{\text{validity}}^{(k)} + R_{\text{intra}}^{(k)} + R_{\text{inter}}^{(k)}. 
\end{equation} 
This reward directly operationalizes our objective: guiding the model to produce a compact set of valid hypotheses that collectively approximate the true future distribution. Subsequently, we compute per-hypothesis advantage from the reward and optimize via gradient ascent on the advantage-weighted log-likelihood.

\subsection{Forecasting}
During the inference phase, we employ a stochastic decoding strategy to explore the future outcome space. For a given input $x$, the policy $\pi_\theta$ generates a response $y_k$ containing a sequence of $M$ hypotheses. To capture the intrinsic uncertainty, we repeat this process to sample $K$ independent responses, thereby constructing an inclusive and diverse hypothesis set. 
\section{Experiments}

\subsection{Experimental Setup}
\paragraph{Datasets.}
We evaluate our models on two real-world datasets, OpenForecast~\cite{openforecast} and OpenEP~\cite{openep}. Our model is trained on OpenForecast and evaluated under two settings: in-domain testing on the same dataset, and out-of-domain evaluation on OpenEP to verify its cross-domain generalization. Moreover, we construct a hard subset in both datasets consisting of test samples mispredicted by GPT-4o-mini across all sampling rounds. To mitigate information leakage, we chronologically repartition OpenForecast, using 09/2023 and 01/2024 as cut-off dates to construct the training, validation, and test splits. More details about datasets are provided in Appendix~\ref{app:datasets}.

\paragraph{Evaluation Metrics.}
We evaluate the inclusiveness and diversity of hypotheses using embedding-based soft matching metrics: \textit{SoftPass@K}, measuring whether any ground-truth hypothesis is hit, short as SP@K, \textit{SoftRecall@K}, measuring the proportion of ground-truth hypotheses recalled, short as SR@K, and \textit{ValidRatio@K}, measuring the ratio of unique and valid hypotheses, short as VR@K, where $K$ denotes the number of sampling rounds. Detailed formulations are provided in Appendix~\ref{app:evaluation}. While embedding-based metrics may introduce marginal noise, they offer a reliable and efficient proxy. Therefore, we also report an LLM-based Pass@K, which leverages GPT-5o-mini\footnote{https://platform.openai.com/docs/models/gpt-5-mini} to directly verify correctness. Notably, for $K=1$, we reported the mean and standard deviation averaged over 16 rounds.

\paragraph{Implementation Details.}
We conduct our experiments on two open-source base models: Qwen2.5-3B-Instruct~\cite{qwen25} and Llama3.2-3B-Instruct\footnote{https://www.llama.com/docs/model-cards-and-prompt-formats/llama3\_2/}. We adopt Qwen3-Embedding-4B~\cite{qwen3embedding} as the embedding model for both training and evaluation.
We benchmark our method against three primary baselines: 1) GPT-4o-mini\footnote{https://platform.openai.com/docs/models/gpt-4o-mini}, serving as the non-finetuned backbone; 2) Standard SFT~\cite{peft}; and 3) Standard GRPO~\cite{deepseekmath}, which is optimized solely via a validity reward. Notably, both the GRPO baseline and our proposed method utilize a two-stage training strategy: an SFT warm-up (3k instances) to enforce format alignment, followed by RL fine-tuning (10k instances). To ensure parameter efficiency, all fine-tuning is implemented via LoRA~\cite{lora}. Regarding generation parameters, we standardize the output to 10 hypotheses per round across 16 sampling rounds. More details regarding training configuration can be found in Appendix~\ref{app:configuration}.

\subsection{Main Results}
\begin{table*}[h]
\centering
\resizebox{\textwidth}{!}{%
\begin{tabular}{lcccc|cccc}
\toprule
\multicolumn{1}{c}{\multirow{2}{*}{\textbf{Model}}} & \multicolumn{4}{c}{\textbf{OpenForecast}} & \multicolumn{4}{c}{\textbf{OpenEP}} \\ \cmidrule{2-9}
\multicolumn{1}{c}{} & \textbf{Pass@1/Pass@16} & \textbf{SP@1/SP@16} & \textbf{SR@1/SR@16} & \textbf{VR@16} & \textbf{Pass@1/Pass@16} & \textbf{SP@1/SP@16} & \textbf{SR@1/SR@16} & \textbf{VR@16} \\ \midrule

\textbf{GPT-4o-mini} 
& $7.55^{\textcolor{darkgreen}{\pm 0.73}}/25.44$
& $9.39^{\textcolor{darkgreen}{\pm 0.80}}/22.81$
& $4.96^{\textcolor{darkgreen}{\pm 0.37}}/12.76$
& $4.23$ 
& $18.82^{\textcolor{darkgreen}{\pm 1.44}}/32.22$
& $6.87^{\textcolor{darkgreen}{\pm 1.06}}/17.78$ 
& $3.06^{\textcolor{darkgreen}{\pm 0.81}}/9.28$ 
& $4.08$ \\ \midrule

\multicolumn{9}{l}{\textit{\textbf{Llama3.2-3B-Instruct}}} \\ \midrule
\textbf{Base} 
& $4.40^{\textcolor{darkgreen}{\pm 0.72}}/21.71$ 
& $9.40^{\textcolor{darkgreen}{\pm 0.94}}/\underline{27.19}$ 
& $4.85^{\textcolor{darkgreen}{\pm 0.53}}/\underline{15.42}$
& $5.73$
& $\mathbf{18.26}^{\textcolor{darkgreen}{\pm 1.76}}/\mathbf{33.33}$ 
& $7.29^{\textcolor{darkgreen}{\pm 1.57}}/18.89$
& $2.97^{\textcolor{darkgreen}{\pm 0.76}}/9.72$
& $\underline{5.67}$ \\

\textbf{+ SFT} 
& $4.50^{\textcolor{darkgreen}{\pm 1.05}}/20.39$ 
& $7.10^{\textcolor{darkgreen}{\pm 0.66}}/21.27$
& $3.64^{\textcolor{darkgreen}{\pm 0.45}}/11.72$
& $\mathbf{7.74}$
& $6.81^{\textcolor{darkgreen}{\pm 2.00}}/25.56$ 
& $9.10^{\textcolor{darkgreen}{\pm 1.85}}/\underline{30.00}$ 
& $2.37^{\textcolor{darkgreen}{\pm 0.54}}/\underline{11.43}$
& $\mathbf{6.92}$ \\

\textbf{+ GRPO} 
& $\mathbf{11.90}^{\textcolor{darkgreen}{\pm 1.05}}/\underline{28.29}$ 
& $\underline{13.72}^{\textcolor{darkgreen}{\pm 0.74}}/21.05$
& $\underline{7.08}^{\textcolor{darkgreen}{\pm 0.42}}/11.61$
& $0.71$
& $2.85^{\textcolor{darkgreen}{\pm 1.30}}/5.56$ 
& $\underline{11.11}^{\textcolor{darkgreen}{\pm 1.52}}/20.00$ 
& $\underline{3.14}^{\textcolor{darkgreen}{\pm 0.70}}/6.35$ 
& $0.62$  \\

\textbf{+ SCATTER} 
& $\underline{7.73}^{\textcolor{darkgreen}{\pm 0.80}}/\mathbf{31.36}$ 
& $\mathbf{17.45}^{\textcolor{darkgreen}{\pm 0.81}}/\mathbf{42.32}$ 
& $\mathbf{9.47}^{\textcolor{darkgreen}{\pm 0.55}}/\mathbf{25.64}$
& $\underline{6.54}$
& $\underline{7.36}^{\textcolor{darkgreen}{\pm 2.29}}/\underline{25.56}$
& $\mathbf{15.69}^{\textcolor{darkgreen}{\pm 2.63}}/\mathbf{36.67}$
& $\mathbf{4.67}^{\textcolor{darkgreen}{\pm 1.03}}/\mathbf{15.30}$ 
& $4.90$ \\ \midrule 

\multicolumn{9}{l}{\textbf{\textit{Qwen2.5-3B-Instruct}}} \\ \midrule
\textbf{Base} 
& $5.96^{\textcolor{darkgreen}{\pm 1.01}}/\underline{25.88}$ 
& $6.95^{\textcolor{darkgreen}{\pm 0.58}}/20.61$ 
& $3.67^{\textcolor{darkgreen}{\pm 0.26}}/10.72$
& $5.07$
& $\mathbf{15.00}^{\textcolor{darkgreen}{\pm 2.22}}/\underline{32.22}$ 
& $4.93^{\textcolor{darkgreen}{\pm 1.71}}/17.78$
& $2.51^{\textcolor{darkgreen}{\pm 0.95}}/10.20$ 
& $\underline{5.56}$  \\

\textbf{+ SFT} 
& $5.51^{\textcolor{darkgreen}{\pm 0.46}}/23.68$
& $8.55^{\textcolor{darkgreen}{\pm 0.91}}/24.56$ 
& $4.48^{\textcolor{darkgreen}{\pm 0.45}}/13.85$
& $\underline{7.10}$ 
& $5.00^{\textcolor{darkgreen}{\pm 2.19}}/20.00$ 
& $6.46^{\textcolor{darkgreen}{\pm 1.72}}/21.11$
& $2.02^{\textcolor{darkgreen}{\pm 0.71}}/8.38$
& $5.52$  \\

\textbf{+ GRPO} 
& $\mathbf{11.84}^{\textcolor{darkgreen}{\pm 0.98}}/25.66$
& $\underline{17.60}^{\textcolor{darkgreen}{\pm 0.66}}/\underline{27.85}$ 
& $\underline{9.70}^{\textcolor{darkgreen}{\pm 0.46}}/\underline{15.94}$ 
& $0.87$
& $5.49^{\textcolor{darkgreen}{\pm 1.73}}/13.33$
& $\underline{19.65}^{\textcolor{darkgreen}{\pm 1.51}}/\underline{26.67}$
& $\mathbf{6.92}^{\textcolor{darkgreen}{\pm 0.53}}/\underline{10.58}$
& $0.76$ \\

\textbf{+ SCATTER} 
& $\underline{8.55}^{\textcolor{darkgreen}{\pm 0.87}}/\mathbf{30.26}$
& $\mathbf{17.61}^{\textcolor{darkgreen}{\pm 1.12}}/\mathbf{41.89}$ 
& $\mathbf{9.77}^{\textcolor{darkgreen}{\pm 0.59}}/\mathbf{24.29}$ 
& $\mathbf{9.14}$ 
& $\underline{9.65}^{\textcolor{darkgreen}{\pm 2.71}}/\mathbf{35.56}$ 
& $\mathbf{16.25}^{\textcolor{darkgreen}{\pm 2.75}}/\mathbf{38.89}$ 
& $\underline{4.88}^{\textcolor{darkgreen}{\pm 0.93}}/\mathbf{17.59}$
& $\mathbf{8.21}$  \\  \bottomrule
\end{tabular}%
}
\vspace{-3pt}
\caption{Main results. Best scores for each backbone are \textbf{bolded} and \underline{underlined} for the second best. The green superscript indicates the standard deviation.}
\vspace{-6pt}
\label{tab:main_result}
\end{table*}

\begin{table*}[h]
\centering
\resizebox{\textwidth}{!}{%
\begin{tabular}{lcccc|cccc}
\toprule
\multicolumn{1}{c}{\multirow{2}{*}{\textbf{Model}}} & \multicolumn{4}{c}{\textbf{OpenForecast-Hard (N=335)}} & \multicolumn{4}{c}{\textbf{OpenEP-Hard (N=61)}} \\ \cmidrule{2-9}
\multicolumn{1}{c}{} & \textbf{Pass@1/Pass@16} & \textbf{SP@1/SP@16} & \textbf{SR@1/SR@16} & \textbf{VR@16} & \textbf{Pass@1/Pass@16} & \textbf{SP@1/SP@16} & \textbf{SR@1/SR@16} & \textbf{VR@16} \\ \midrule

\textbf{GPT-4o-mini} 
& $0.00^{\textcolor{darkgreen}{\pm 0.00}}/0.00$  
& $4.65^{\textcolor{darkgreen}{\pm 0.59}}/14.93$
& $1.76^{\textcolor{darkgreen}{\pm 0.35}}/6.75$
& $4.10$ 
& $0.00^{\textcolor{darkgreen}{\pm 0.00}}/0.00$   
& $2.56^{\textcolor{darkgreen}{\pm 1.29}}/4.92$
& $0.92^{\textcolor{darkgreen}{\pm 0.50}}/3.28$  
& $4.03$ \\ \midrule

\multicolumn{9}{l}{\textit{\textbf{Llama3.2-3B-Instruct}}} \\ \midrule
\textbf{Base} 
& $0.95^{\textcolor{darkgreen}{\pm 0.55}}/10.15$
& $6.16^{\textcolor{darkgreen}{\pm 0.97}}/\underline{20.00}$ 
& $2.55^{\textcolor{darkgreen}{\pm 0.43}}/\underline{9.43}$
& $5.56$
& $1.23^{\textcolor{darkgreen}{\pm 1.08}}/9.84$
& $3.69^{\textcolor{darkgreen}{\pm 1.36}}/9.84$
& $1.27^{\textcolor{darkgreen}{\pm 0.47}}/3.83$
& $\underline{5.75}$ \\

\textbf{+ SFT} 
& $2.05^{\textcolor{darkgreen}{\pm 0.71}}/13.13$
& $4.79^{\textcolor{darkgreen}{\pm 0.90}}/16.72$ 
& $1.86^{\textcolor{darkgreen}{\pm 0.45}}/7.70$  
& $\mathbf{7.71}$
& $\underline{4.41}^{\textcolor{darkgreen}{\pm 2.37}}/\underline{16.39}$
& $7.07^{\textcolor{darkgreen}{\pm 2.51}}/\underline{24.59}$
& $1.86^{\textcolor{darkgreen}{\pm 0.79}}/\underline{8.69}$
& $\mathbf{6.68}$ \\

\textbf{+ GRPO} 
& $\mathbf{6.55}^{\textcolor{darkgreen}{\pm 0.98}}/\underline{21.19}$
& $\underline{9.27}^{\textcolor{darkgreen}{\pm 0.54}}/15.22$
& $\underline{3.67}^{\textcolor{darkgreen}{\pm 0.35}}/6.99$
& $0.75$
& $0.51^{\textcolor{darkgreen}{\pm 0.76}}/1.64$
& $\underline{11.27}^{\textcolor{darkgreen}{\pm 1.52}}/19.67$
& $\underline{3.05}^{\textcolor{darkgreen}{\pm 0.39}}/5.23$
& $0.62$ \\

\textbf{+ SCATTER} 
& $\underline{3.75}^{\textcolor{darkgreen}{\pm 0.75}}/\mathbf{23.28}$
& $\mathbf{13.15}^{\textcolor{darkgreen}{\pm 0.96}}/\mathbf{34.33}$
& $\mathbf{6.14}^{\textcolor{darkgreen}{\pm 0.65}}/\mathbf{19.09}$
& $\underline{6.58}$
& $\mathbf{5.94}^{\textcolor{darkgreen}{\pm 2.24}}/\mathbf{16.39}$
& $\mathbf{15.27}^{\textcolor{darkgreen}{\pm 2.37}}/\mathbf{27.87}$
& $\mathbf{4.65}^{\textcolor{darkgreen}{\pm 0.76}}/\mathbf{11.44}$
& $4.71$ \\ \midrule

\multicolumn{9}{l}{\textit{\textbf{Qwen2.5-3B-Instruct}}} \\ \midrule
\textbf{Base} 
& $1.47^{\textcolor{darkgreen}{\pm 0.43}}/10.75$
& $4.68^{\textcolor{darkgreen}{\pm 0.54}}/14.63$
& $1.71^{\textcolor{darkgreen}{\pm 0.31}}/5.74$
& $4.93$
& $\underline{2.56}^{\textcolor{darkgreen}{\pm 1.42}}/\underline{6.56}$
& $1.84^{\textcolor{darkgreen}{\pm 1.82}}/9.84$
& $0.85^{\textcolor{darkgreen}{\pm 0.80}}/5.33$ 
& $5.13$ \\

\textbf{+ SFT} 
& $2.16^{\textcolor{darkgreen}{\pm 0.43}}/13.43$
& $5.45^{\textcolor{darkgreen}{\pm 0.88}}/17.61$
& $2.35^{\textcolor{darkgreen}{\pm 0.53}}/8.95$
& $\underline{6.99}$
& $1.23^{\textcolor{darkgreen}{\pm 1.08}}/4.92$
& $4.41^{\textcolor{darkgreen}{\pm 1.50}}/14.75$ 
& $1.37^{\textcolor{darkgreen}{\pm 0.51}}/4.92$
& $\underline{5.44}$ \\

\textbf{+ GRPO} 
& $\mathbf{6.36}^{\textcolor{darkgreen}{\pm 0.77}}/\underline{17.01}$
& $\underline{12.95}^{\textcolor{darkgreen}{\pm 0.85}}/\underline{21.19}$
& $\underline{5.68}^{\textcolor{darkgreen}{\pm 0.50}}/\underline{10.56}$ 
& $0.87$
& $2.15^{\textcolor{darkgreen}{\pm 1.61}}/6.56$
& $\underline{16.60}^{\textcolor{darkgreen}{\pm 1.73}}/\underline{21.31}$ 
& $\underline{4.99}^{\textcolor{darkgreen}{\pm 0.49}}/\underline{7.91}$
& $0.74$ \\

\textbf{+ SCATTER} 
& $\underline{4.12}^{\textcolor{darkgreen}{\pm 0.79}}/\mathbf{22.39}$
& $\mathbf{13.40}^{\textcolor{darkgreen}{\pm 1.07}}/\mathbf{34.93}$ 
& $\mathbf{6.54}^{\textcolor{darkgreen}{\pm 0.62}}/\mathbf{17.92}$
& $\mathbf{8.99}$
& $\mathbf{6.66}^{\textcolor{darkgreen}{\pm 2.05}}/\mathbf{21.31}$
& $13.42^{\textcolor{darkgreen}{\pm 3.90}}/\mathbf{31.15}$ 
& $4.39^{\textcolor{darkgreen}{\pm 1.18}}/\mathbf{15.44}$
& $\mathbf{8.15}$ \\ \bottomrule
\end{tabular}%
}
\caption{Performance on the subsets of hard samples. Best scores for each backbone are \textbf{bolded}, and second best are \underline{underlined}. The green superscript indicates the standard deviation.}
\vspace{-6pt}
\label{tab:hard_result}
\end{table*}

\paragraph{Overall Performance.}
Table~\ref{tab:main_result} presents the overall performance of SCATTER against baseline methods (Base, SFT) and the standard RL baseline (GRPO) across two distinct backbones. SCATTER equipped with Qwen2.5-3B-Instruct achieves state-of-the-art results. This performance indicates that our hybrid reward mechanism successfully promotes the inclusiveness and diversity of the generated hypotheses. To rigorously assess robustness against more complex scenarios, we further evaluate performance on a curated hard subset, defined as the specific collection of test samples where GPT-4o-mini to predict the correct outcome. More detailed information is provided in Appendix~\ref{app:datasets}. As detailed in Table~\ref{tab:hard_result}, our method maintains a significant performance advantage on these challenging cases, demonstrating enhanced effectiveness even when facing tasks of increased difficulty.
\begin{table}[h]
\centering
\resizebox{\linewidth}{!}{%
\begin{tabular}{lcccc|cccc}
\toprule
\multirow{2}{*}{\textbf{Model}} & \multicolumn{4}{c}{\textbf{OpenForecast}} & \multicolumn{4}{c}{\textbf{OpenEP}} \\
 \cmidrule(lr){2-9}
 & \textbf{Pass} & \textbf{SP} & \textbf{SR} & \textbf{VR} & \textbf{Pass} & \textbf{SP} & \textbf{SR} & \textbf{VR} \\ 
\midrule
\multicolumn{9}{l}{\textit{\textbf{Llama3.2-3B-Instruct}}} \\ \midrule
\textbf{SCATTER} & 31.36 & 42.32 & 25.64 & \textbf{6.54} & \textbf{25.56} & \textbf{36.67} & \textbf{15.30} & \textbf{4.90} \\
\quad \textit{-inter} & \textbf{34.43} & \textbf{44.74} & \textbf{25.91} & 4.36 & 22.22 & 33.33 & 14.42 & 3.97 \\
\quad \textit{-intra} & 31.14 & 42.11 & 25.36 & 2.64 & 23.33 & 30.00 & 12.45 & 1.93 \\
\textbf{GRPO} & 28.29 & 21.05 & 11.61 & 0.71 & 5.56 & 20.00 & 6.35 & 0.62 \\
\midrule
\multicolumn{9}{l}{\textit{\textbf{Qwen2.5-3B-Instruct}}} \\ \midrule
\textbf{SCATTER} & 30.26 & \textbf{41.89} & 24.29 & \textbf{9.14} & \textbf{35.56} & 38.89 & 17.59 & \textbf{8.21} \\
\quad \textit{-inter} & \textbf{33.11} & 41.01 & \textbf{24.44} & 4.74 & 33.33 & \textbf{40.00} & \textbf{18.84} & 4.78 \\
\quad \textit{-intra} & 33.11 & 41.45 & 24.14 & 4.19 & 32.22 & 40.0 & 18.7 & 4.14 \\ 
\textbf{GRPO} & 25.66 & 27.85 & 15.94 & 0.87 & 13.33 & 26.67 & 10.58 & 0.76 \\
\bottomrule
\end{tabular}%
}
\caption{Performance comparison \wrt various reward designs. Pass, SP, SR, and VR are based on $K=16$.}
\vspace{-10pt}
\label{tab:reward_design}
\end{table}
On the standard dataset, the base model achieves performance comparable to GPT-4o-mini but exhibits a significantly higher validity ratio. Surprisingly, on the hard subset, the base model even outperforms GPT-4o-mini. We attribute this anomaly to the alignment tax~\cite{instruct_training,alignmenttax} inherent in safety-aligned commercial models; while GPT-4o-mini has stronger reasoning capabilities, extensive pre-training and post-training alignment bias the model towards high-probability events.
It is noteworthy that the standard GRPO baseline occasionally achieves higher \textit{Pass@1} scores. However, this comes at a severe cost: GRPO suffers from mode collapse, with \textit{ValidRatio@16} dropping to near-zero levels. This observation suggests that without explicit validity constraints, standard GRPO exploits the reward signal by aggressively optimizing for inclusiveness, producing "high-scoring" but semantically repetitive and linguistically broken outputs.

\paragraph{Domain Generalization.} Notably, SCATTER maintains robust performance even under out-of-domain settings, suggesting superior generalization capabilities under distribution shifts. We attribute this to our explicit optimization objective, which effectively balances hypothesis diversity with semantic validity. Interestingly, we observe that post-training methods on Llama3.2-3B-Instruct exhibit a performance regression compared to the backbone on standard subsets. We posit that Llama-3.2 exhibits a heightened sensitivity to the exploration noise inherent in reinforcement learning, which subsequently undermines its instruction-following stability. This is empirically supported by the significantly lower \textit{ValidRatio@16} observed in Llama-based experiments, suggesting that the propensity for generating invalid responses increases when incentivized to explore a diverse output space.

\begin{table}[h]
\centering
\resizebox{\linewidth}{!}{%
\begin{tabular}{lcccc|cccc}
\toprule
\multirow{2}{*}{\textbf{Model}} & \multicolumn{4}{c}{\textbf{OpenForecast}} & \multicolumn{4}{c}{\textbf{OpenEP}} \\
 \cmidrule(lr){2-9}
 & \textbf{Pass} & \textbf{SP} & \textbf{SR} & \textbf{VR} & \textbf{Pass} & \textbf{SP} & \textbf{SR} & \textbf{VR} \\ 
\midrule
\textbf{Vanilla} & 19.3 & 24.56 & 13.84 & \textbf{17.77}  & 11.11 & 15.56 & 6.9 & \textbf{15.7} \\
\textbf{Mean} & 11.09 & 41.24 & 24.08 & 9.54 &  26.67 & 27.78 & 11.59 & 8.49 \\
\textbf{Min} & \textbf{31.80} & 41.67 & 23.90 & 8.21 & 18.89 & 23.33 & 9.41 & 7.17 \\
\textbf{SCATTER} & 30.26 & \textbf{41.89} & \textbf{24.29} & 9.14 & \textbf{35.56} & \textbf{38.89} & \textbf{17.59} & 8.21 \\
\bottomrule
\end{tabular}%
}
\caption{Performance comparison \wrt various validity gated score. Pass, SP, SR, and VR are based on $K=16$.}
\vspace{-10pt}
\label{tab:validity_score}
\end{table}

\subsection{Ablation Study}

\paragraph{Reward Design.} To investigate the individual contributions of our proposed components, we compare SCATTER against two variants: 1) \textit{-inter}, which removes the inter-group diversity reward, and 2) \textit{-intra}, which excludes the intra-group diversity constraint. Table~\ref{tab:reward_design} summarizes the results.
We observe that the full SCATTER framework achieves the most robust balance between performance and generalization. First, regarding the intra-group diversity component, its removal leads to a catastrophic drop in response validity, particularly in the out-of-domain setting. This indicates that enforcing local diversity is essential for maintaining the semantic quality of hypotheses. Second, while the \textit{-inter} variant occasionally exhibits higher pass scores on the in-domain dataset, this performance is deceptive. It comes at the cost of significantly lower validity and, crucially, reduced generalization on the out-of-domain benchmark. This suggests that the inter-group diversity acts as a global regularizer, preventing the model from overfitting to in-domain patterns with high-entropy but low-validity outputs, thereby ensuring robust adaptation to unseen domains.

\begin{figure}[t]
  \includegraphics[width=\columnwidth]{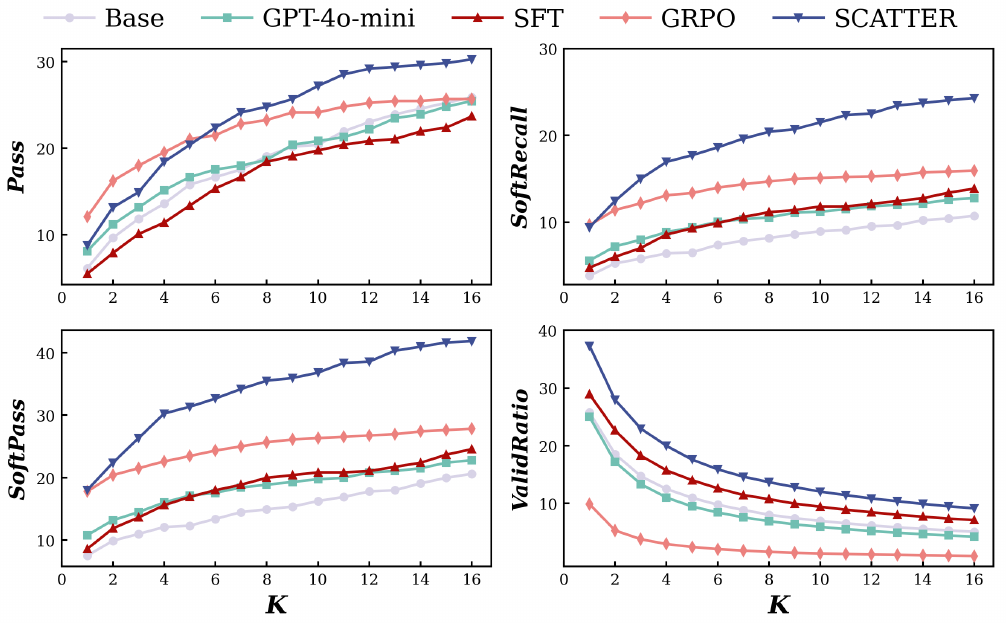}
  \caption{Performance scaling \wrt sampling rounds $K$. Comparison of SCATTER against baselines on OpenForecast using Qwen2.5-3B-Instruct as Base model. }
  \vspace{-10pt}
  \label{fig:sampling}
\end{figure}
\paragraph{Validity Gated Score.} We conduct an ablation study to isolate the contribution of the validity aggregation strategy. Specifically, we assess whether the validity gate mechanism improves the ability to balance inclusiveness and diversity. We compare the full model against three variants: \textbf{Vanilla}, which excludes the validity-gated score; \textbf{Mean}, which utilizes the mean of the scores; and \textbf{Min}, which adopts the minimum score. As shown in Table~\ref{tab:validity_score}, the complete SCATTER method demonstrates the most robust overall performance. On OpenEP, it achieves state-of-the-art results across all accuracy metrics. Similarly, on OpenForecast, SCATTER leads in both \textit{SoftPass@16} and \textit{SoftRecall@16}. In contrast, while Vanilla attains the highest validity ratio, its significantly lower performance on accuracy-oriented metrics indicates that optimizing for validity alone, without effective aggregation, compromises the correctness of the generated hypotheses. Although SCATTER-Min shows competitive performance on OpenForecast, the full SCATTER framework provides a superior balance between accuracy and diversity.

\subsection{Model Study}
\label{model_study}
\paragraph{Impact of Number of Sampling Round $K$.}
Figure~\ref{fig:sampling} illustrates the performance scaling of Qwen2.5 across varying sampling rounds ($K \in [1, 16]$). Additional results regarding Llama3.2 can be found in Appendix~\ref{app:sampling_llama}. In terms of inclusiveness, SCATTER demonstrates superior scaling properties. Unlike GRPO, which saturates at high $K$, SCATTER scales effectively and significantly outperforms backbone and SFT models on \textit{SoftPass@k}. Crucially, this gain in diversity does not come at the expense of quality; while the \textit{ValidRatio@k} for GRPO suffers a precipitous decline toward near-zero values, SCATTER consistently retains the highest validity density. This confirms that our method effectively leverages expanded compute budgets to uncover a broader spectrum of semantically valid hypotheses without drifting into plausible-but-incorrect generation modes.

\begin{figure}[t]
  \includegraphics[width=\columnwidth]{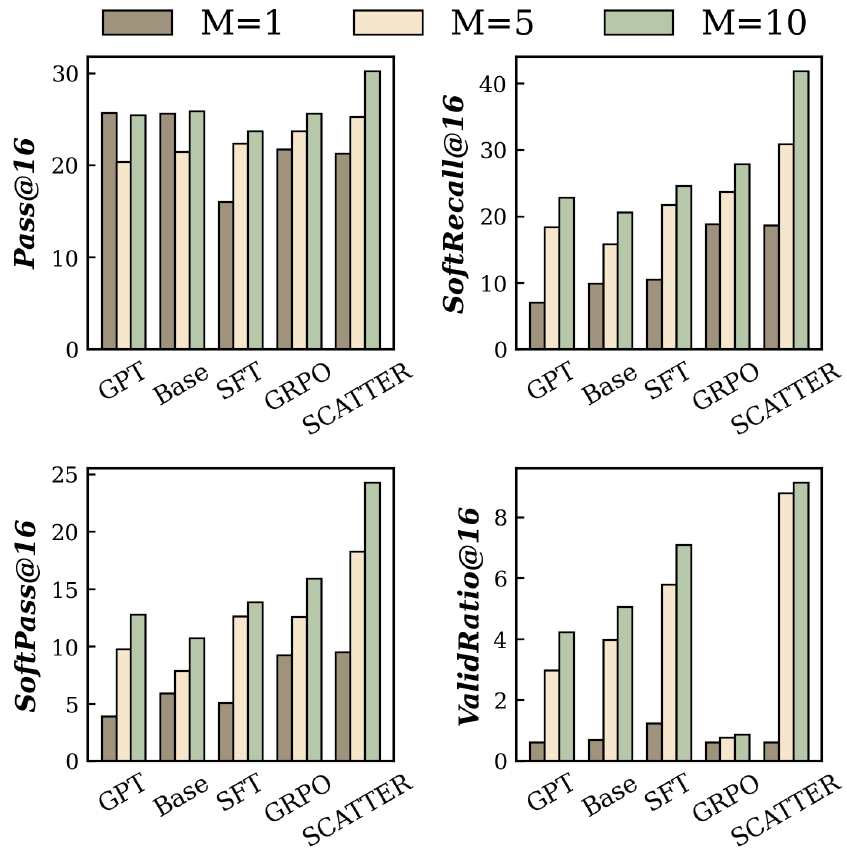}
  \caption{Performance scaling \wrt the number of hypotheses $M$ per round.}
  \label{fig:m_scaling}
    \vspace{-10pt}
\end{figure}

\paragraph{Impact of Number of Hypothesis $M$.} Figure~\ref{fig:m_scaling} illustrates the impact of varying $M \in \{1, 5, 10\}$ on model performance. SCATTER exhibits the most robust scaling properties among all compared methods. SCATTER maintains a steep upward trajectory. Notably, at $M=10$, SCATTER achieves significant margins over the strongest baseline. This confirms that our method can effectively leverage larger $M$ values to enhance semantic coverage and accuracy. However, baselines such as GPT and the base model exhibit limited improvement or saturation as $M$ increases. Notably, a performance dip occurs when transitioning from deterministic ($M=1$) to stochastic decoding. This decline is attributed to the introduction of noise and local optima, which small sample sizes fail to mitigate until a sufficient number of hypotheses are generated.

\paragraph{Case Study.}
\begin{figure}[t]
  \includegraphics[width=\columnwidth]{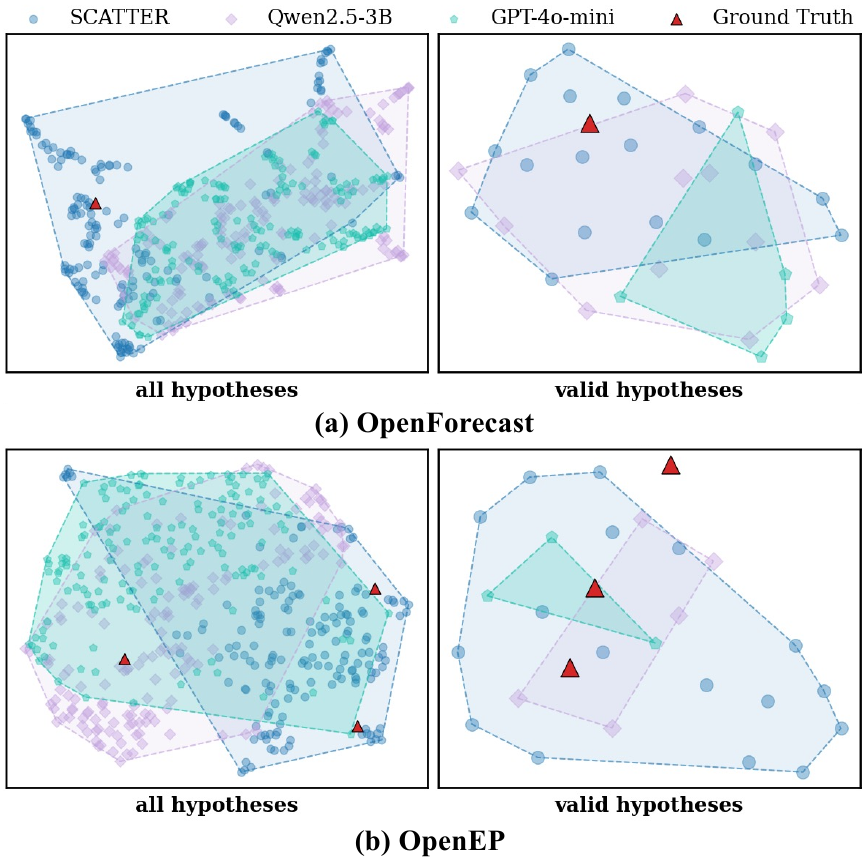}
  \caption{UMAP visualization of the semantic manifold of the generated hypotheses on two datasets. }
  \vspace{-10pt}
  \label{fig:case_study}
\end{figure}
To provide a qualitative perspective on the mechanism behind SCATTER’s superior performance, Figure~\ref{fig:case_study} visualizes the semantic manifold of generated hypotheses via UMAP. SCATTER demonstrates a robust balance between semantic exploration and ground truth alignment. In the distribution of all hypotheses, SCATTER shows significantly broader coverage than baselines. More importantly, the distribution of valid hypotheses indicates that this exploration is constructive. While SCATTER shows broader coverage overall, the significantly more uniform distribution of its valid hypotheses is particularly notable. This indicates constructive exploration, demonstrating that our method effectively confines exploration to contextually plausible futures. Specific examples of generated hypotheses are provided in the Appendix~\ref{app:case_study}.
\section{Conclusion}
In this paper, we address the limitations of the prevailing single-prediction paradigm in open-ended event forecasting and advocate for a shift toward \textit{scatter forecasting}. By reformulating the task as hypothesis generation, we aim to better capture the intrinsic uncertainty of real-world developments. To realize this, we propose SCATTER, a reinforcement learning framework designed to simultaneously optimize the inclusiveness and diversity of generated hypotheses. 
Extensive experiments on the OpenForecast and OpenEP benchmarks demonstrate that our approach significantly outperforms existing baselines, producing hypotheses that are both diverse and high-quality. 

\section*{Limitations}
In this work, we identify several limitations to be addressed in future research. First, due to computational constraints and API costs, our experiments are primarily conducted on 3B-parameter models using LoRA fine-tuning, and our evaluation of closed-source commercial LLMs is restricted to GPT-4o-mini. Second, we set the maximum number of sampling rounds ($K$) to 16. Given the inherent uncertainty of LLMs, minor statistical fluctuations may occur across different runs; however, these marginal deviations do not compromise the validity of our core experimental conclusions. Finally, although the datasets  we adopt contain multiple ground-truth references, they still capture a limited subset of the complexity found in real-world events. To address this gap, we will extend our framework to label-free settings in the future work. Additionally, AI assistants were used solely for language editing and polishing, without affecting the research content, methodology, or conclusions.
\section{Acknowldegements}

We thank the anonymous reviewers for their valuable comments. This research was supported by the Singapore Ministry of Education (MOE) Academic Research Fund (AcRF) Tier 1 grant.
\bibliography{custom}

\appendix
\section{Appendix}
\label{sec:appendix}

\subsection{Datasets Details}
\label{app:datasets}
We evaluate our method using two real-world datasets: OpenEP~\cite{openep} and OpenForecast~\cite{openforecast}. For OpenEP, since each answer corresponds to a span in a real news article, we re-extract answer spans to ensure evaluation reliability. Refer to Appendix~\ref{app:prompt} for the specific prompts utilized. Notably, we filter out time and location prediction queries due to their high uncertainty and task misalignment. For OpenForecast, we focus on the short-term forecasting subset, which features precise timestamps and allows for multiple valid answers.
o mitigate information leakage, we re-partitioned the OpenForecast dataset based on the knowledge cutoff date of the employed LLM. The knowledge cutoff dates are listed in Table~\ref{tab:cutoff}, and detailed statistics for both datasets are provided in Table~\ref{tab:dataset_stats}.

\begin{table}[t]
\centering
\resizebox{\linewidth}{!}{%
\begin{tabular}{lccc}
\toprule
\textbf{Dataset} & \textbf{Split} & \textbf{\#Questions} & \textbf{Time Span} \\
\midrule
\multirow{3}{*}{\textbf{OpenForecast}}
& Train & 32596 & 1950-02 $\sim$ 2023-08 \\
& Valid & 654   & 2023-09 $\sim$ 2023-12 \\
& Test  & 456   & 2024-01 $\sim$ 2024-12 \\
& Hard-Test  & 335   & 2024-01 $\sim$ 2024-12 \\
\midrule
\multirow{2}{*}{\textbf{OpenEP}}
& Test  & 90    & 2024-07 $\sim$  \\
& Hard-Test  & 61    & 2024-07 $\sim$  \\
\bottomrule
\end{tabular}}
\caption{Dataset statistics for OpenForecast and OpenEP. \#Questions denotes the number of samples.}
\label{tab:dataset_stats}
\end{table}

\begin{table}[t]
    \centering
    \small
    \begin{tabular}{lc} 
        \toprule
        \textbf{Model} & \textbf{Knowledge Cutoff} \\
        \midrule
        GPT-4o mini & Oct. 2023 \\
        Llama-3.2-3B-Instruct & Dec. 2023 \\
        Qwen2.5-3B-Instruct & Dec. 2023 \\
        \bottomrule
    \end{tabular}
    \caption{Knowledge cutoff dates of the employed LLMs.}
    \label{tab:cutoff} 
\end{table}

\begin{table*}[h]
\centering
\resizebox{0.9\textwidth}{!}{%
\begin{tabular}{lccc|ccc}
\toprule
\multicolumn{1}{c}{\multirow{2}{*}{\textbf{Model}}} & \multicolumn{3}{c}{\textbf{OpenForecast}} & \multicolumn{3}{c}{\textbf{OpenEP}} \\ \cmidrule{2-7}
\multicolumn{1}{c}{} & \textbf{SP@1/SP@16} & \textbf{SR@1/SR@16} & \textbf{VR@16}  & \textbf{SP@1/SP@16} & \textbf{SR@1/SR@16} & \textbf{VR@16} \\ \midrule

\textbf{Base} 
& $1.43^{\textcolor{darkgreen}{\pm 0.36}}/6.14$ 
& $0.82^{\textcolor{darkgreen}{\pm 0.28}}/3.81$
& $24.45$
& $0.49^{\textcolor{darkgreen}{\pm 0.68}}/3.33$
& $0.15^{\textcolor{darkgreen}{\pm 0.24}}/1.00$
& $25.69$ \\

\textbf{+ SFT} 
& $2.12^{\textcolor{darkgreen}{\pm 0.39}}/8.77$
& $1.14^{\textcolor{darkgreen}{\pm 0.24}}/4.42$
& $22.03$
& $1.04^{\textcolor{darkgreen}{\pm 0.92}}/6.67$ 
& $0.33^{\textcolor{darkgreen}{\pm 0.40}}/2.75$
& $19.38$ \\

\textbf{+ GRPO} 
& $6.35^{\textcolor{darkgreen}{\pm 0.38}}/11.62$
& $3.33^{\textcolor{darkgreen}{\pm 0.21}}/5.94$
& $2.39$
& $6.88^{\textcolor{darkgreen}{\pm 1.53}}/15.56$ 
& $1.86^{\textcolor{darkgreen}{\pm 0.35}}/4.41$ 
& $2.26$  \\

\textbf{+ SCATTER} 
& $6.21^{\textcolor{darkgreen}{\pm 0.63}}/18.64$ 
& $3.38^{\textcolor{darkgreen}{\pm 0.40}}/9.85$
& $28.41$
& $4.38^{\textcolor{darkgreen}{\pm 1.64}}/17.78$
& $1.02^{\textcolor{darkgreen}{\pm 0.39}}/5.29$ 
& $26.01$ \\  \bottomrule
\end{tabular}%
}
\vspace{-3pt}
\caption{Cross-embedding evaluation results of Qwen2.5-3B-Instruct using all-MiniLM-L6-v2. The green superscript indicates the standard deviation.}
\vspace{-6pt}
\label{tab:cross_embedding_minilm}
\end{table*}

\begin{table*}[h]
\centering
\resizebox{0.9\textwidth}{!}{%
\begin{tabular}{lccc|ccc}
\toprule
\multicolumn{1}{c}{\multirow{2}{*}{\textbf{Model}}} & \multicolumn{3}{c}{\textbf{OpenForecast}} & \multicolumn{3}{c}{\textbf{OpenEP}} \\ \cmidrule{2-7}
\multicolumn{1}{c}{} & \textbf{SP@1/SP@16} & \textbf{SR@1/SR@16} & \textbf{VR@16}  & \textbf{SP@1/SP@16} & \textbf{SR@1/SR@16} & \textbf{VR@16} \\ \midrule

\textbf{Base} 
& $23.68^{\textcolor{darkgreen}{\pm 0.93}}/46.71$ 
& $11.63^{\textcolor{darkgreen}{\pm 0.69}}/26.40$
& $1.04$
& $35.14^{\textcolor{darkgreen}{\pm 3.64}}/67.78$
& $14.60^{\textcolor{darkgreen}{\pm 1.66}}/37.72$
& $1.28$ \\

\textbf{+ SFT} 
& $23.73^{\textcolor{darkgreen}{\pm 0.69}}/52.19$
& $11.39^{\textcolor{darkgreen}{\pm 0.48}}/30.00$
& $2.08$
& $30.49^{\textcolor{darkgreen}{\pm 3.45}}/70.00$ 
& $13.49^{\textcolor{darkgreen}{\pm 1.61}}/39.98$
& $1.56$ \\

\textbf{+ GRPO} 
& $23.99^{\textcolor{darkgreen}{\pm 1.02}}/41.01$
& $12.24^{\textcolor{darkgreen}{\pm 0.66}}/24.08$
& $0.73$
& $26.74^{\textcolor{darkgreen}{\pm 2.47}}/46.67$ 
& $11.16^{\textcolor{darkgreen}{\pm 1.07}}/21.37$ 
& $0.71$  \\

\textbf{+ SCATTER} 
& $32.11^{\textcolor{darkgreen}{\pm 1.25}}/64.04$ 
& $16.63^{\textcolor{darkgreen}{\pm 0.76}}/41.87$
& $3.18$
& $40.35^{\textcolor{darkgreen}{\pm 3.53}}/73.33$
& $19.97^{\textcolor{darkgreen}{\pm 2.08}}/42.10$ 
& $3.28$ \\   \bottomrule
\end{tabular}%
}
\vspace{-3pt}
\caption{Cross-embedding evaluation results of Qwen2.5-3B-Instruct using gemma-300m. The green superscript indicates the standard deviation.}
\vspace{-6pt}
\label{tab:cross_embedding_gemma}
\end{table*}

\subsection{Evaluation Details}
\label{app:evaluation}

To rigorously assess the inclusiveness and diversity of the generated hypotheses, we employ a hybrid evaluation framework combining embedding-based metrics with LLM-based verification. We define the dataset-level metric as the average of sample-level scores over $N$ test samples: 
\begin{equation}
    \text{Metric}@K = \frac{1}{N} \sum_{n=1}^{N} \text{Score}_n.
\end{equation}
The specific definitions for each metric are detailed below.

\paragraph{SoftPass@K.} This metric measures the generation success rate, indicating whether the model produces \textit{at least one} hypothesis that meets the semantic requirement. Formally, we define the sample-level score as:
\begin{equation}
\text{SoftPass}_n = \mathbb{I}\left( \sum_{k=1}^{K} \sum_{i=1}^{M} \mathbb{I}\left( S_{k,i} > \tau_{sp} \right) \ge 1 \right),
\end{equation}
where $\tau_{sp} = 0.8$ is the similarity threshold, $\mathbb{I}(\cdot)$ is the indicator function. Here, $S_{k,i}=\max_{g \in \mathcal{G}x}cos(\mathbf{h}_{k,i},\mathbf{g})$ is the maximum similarity between the $i$-th hypothesis in the $k$-th round and any ground truth $g \in \mathcal{G}_x$

\paragraph{SoftRecall@K.} SoftRecall assesses the semantic coverage of the ground truth space. It quantifies the proportion of ground truth embeddings that are successfully retrieved by the generated hypothesis set. The score is defined as:
\begin{equation}
\text{SoftRecall}_n = \frac{1}{|\mathcal{G}x|} \sum_{g \in \mathcal{G}x} \mathbb{I}\left( S_{h,g} > \tau_{sr} \right),
\end{equation}
where $\tau_{sr} = 0.8$ serves as the semantic similarity threshold for a successful retrieval and $S_{h,g}$ denote the cosine similarity between the generated hypothesis and the ground truth.
\begin{table*}[h]
\centering
\resizebox{0.77\textwidth}{!}{%
\begin{tabular}{l cc|cc}
\toprule
\multirow{2}{*}{\textbf{Method}} 
& \multicolumn{2}{c}{\textbf{Qwen3-Embedding-4B}} 
& \multicolumn{2}{c}{\textbf{all-MiniLM-L6-v2}} \\ \cmidrule{2-5}
& \textbf{SP@1/SP@16} & \textbf{SR@1/SR@16} 
& \textbf{SP@1/SP@16} & \textbf{SR@1/SR@16} \\ 
\midrule

\multicolumn{5}{l}{\textit{Threshold = 0.70}} \\
\midrule
\textbf{GRPO} 
& $44.78^{\textcolor{darkgreen}{\pm 0.95}}/59.43$
& $27.75^{\textcolor{darkgreen}{\pm 0.59}}/38.60$
& $16.13^{\textcolor{darkgreen}{\pm 0.87}}/28.51$
& $8.37^{\textcolor{darkgreen}{\pm 0.60}}/15.50$ \\

\textbf{SCATTER} 
& $47.81^{\textcolor{darkgreen}{\pm 1.15}}/78.51$
& $29.84^{\textcolor{darkgreen}{\pm 0.80}}/56.79$
& $18.00^{\textcolor{darkgreen}{\pm 0.95}}/46.05$
& $9.74^{\textcolor{darkgreen}{\pm 0.56}}/26.61$ \\

\midrule

\multicolumn{5}{l}{\textit{Threshold = 0.80}} \\
\midrule
\textbf{GRPO} 
& $11.84^{\textcolor{darkgreen}{\pm 0.98}}/25.66$
& $17.60^{\textcolor{darkgreen}{\pm 0.66}}/27.85$
& $6.35^{\textcolor{darkgreen}{\pm 0.38}}/11.62$
& $3.33^{\textcolor{darkgreen}{\pm 0.21}}/5.94$ \\

\textbf{SCATTER} 
& $8.55^{\textcolor{darkgreen}{\pm 0.87}}/30.26$
& $17.61^{\textcolor{darkgreen}{\pm 1.12}}/41.89$
& $6.21^{\textcolor{darkgreen}{\pm 0.63}}/18.64$
& $3.38^{\textcolor{darkgreen}{\pm 0.40}}/9.85$ \\

\midrule

\multicolumn{5}{l}{\textit{Threshold = 0.90}} \\
\midrule
\textbf{GRPO} 
& $3.29^{\textcolor{darkgreen}{\pm 0.45}}/6.36$
& $1.57^{\textcolor{darkgreen}{\pm 0.18}}/2.79$
& $1.40^{\textcolor{darkgreen}{\pm 0.36}}/2.19$
& $0.89^{\textcolor{darkgreen}{\pm 0.24}}/1.35$ \\

\textbf{SCATTER} 
& $2.73^{\textcolor{darkgreen}{\pm 0.38}}/9.87$
& $1.32^{\textcolor{darkgreen}{\pm 0.19}}/4.78$
& $0.66^{\textcolor{darkgreen}{\pm 0.33}}/4.82$
& $0.34^{\textcolor{darkgreen}{\pm 0.18}}/2.57$ \\

\bottomrule
\end{tabular}%
}
\vspace{-3pt}
\caption{Impact of similarity thresholds on SP@16 and SR@16 for OpenForecast. The green superscript indicates the standard deviation.}
\vspace{-6pt}
\label{tab:sp_sr_threshold}
\end{table*}

\begin{table}[h]
\centering
\small
\setlength{\tabcolsep}{5pt}
\begin{tabular}{l cc}
\toprule
\textbf{Method} 
& \textbf{Qwen3-Embedding-4B} 
& \textbf{all-MiniLM-L6-v2} \\
\midrule

\multicolumn{3}{l}{\textit{Threshold = 0.40}} \\
\midrule
\textbf{GRPO}    & 0.66 & 1.05 \\
\textbf{SCATTER} & 2.20 & 9.85 \\

\midrule

\multicolumn{3}{l}{\textit{Threshold = 0.50}} \\
\midrule
\textbf{GRPO}    & 0.72 & 1.50 \\
\textbf{SCATTER} & 5.15 & 17.20 \\

\midrule

\multicolumn{3}{l}{\textit{Threshold = 0.60}} \\
\midrule
\textbf{GRPO}    & 0.87 & 2.39 \\
\textbf{SCATTER} & 9.14 & 28.41 \\

\bottomrule
\end{tabular}
\vspace{-3pt}
\caption{Impact of similarity thresholds on VR@16 for OpenForecast.}
\vspace{-6pt}
\label{tab:vr_threshold}
\end{table}
\paragraph{ValidRatio@K.}To quantify diversity, this metric measures the proportion of valid, non-redundant hypotheses. The score $\text{ValidRatio}_n$ is defined as:\begin{equation}\text{ValidRatio}_n = \frac{1}{K \cdot M} \sum_{k=1}^{K} \sum_{i=1}^{M} V_{k,i},\end{equation}
where $V_{k,i} \in \{0, 1\}$ serves as the validity indicator determined by our diversity filtering mechanism. The indicator $V_{k,i}$ is computed dynamically to enforce distinctiveness: for the initial round ($i=1$), we apply greedy intra-round filtering; for subsequent rounds ($i>1$), candidates are filtered against the history of all valid hypotheses from previous rounds. Additionally, we enforce a minimum semantic relevance by requiring $S_{k,i} \ge \tau_{valid}$, where $\tau_{valid}=0.4$.\paragraph{Pass@K.}Complementing the embedding-based metrics with strict verification, we utilize GPT-5o-mini as an external oracle. Pass@K is defined as the proportion of problems where at least one hypothesis is verified as logically correct by the LLM. The detailed evaluation prompt is provided in Appendix~\ref{app:prompt}.

\subsection{Cross-embedding Evaluation}
To evaluate performance across different embedding models, we use two alternative embedding models, all-MiniLM-L6-v2~\cite{sbert} and gemma-embedding~\cite{gemma}, which are widely adopted and architecturally distinct from the embedding model used for reward computation. Specifically, all-MiniLM-L6-v2 is a lightweight Sentence-BERT model optimized for general semantic similarity, while gemma-embedding is built upon the Gemma foundation model family, providing a different embedding architecture and training paradigm. As reported in Table~\ref{tab:cross_embedding_minilm} and Table~\ref{tab:cross_embedding_gemma}, SCATTER consistently outperforms GRPO and other baselines across both alternative embedding spaces. Notably, the relative performance margins remain stable despite the change of evaluator, demonstrating that the effectiveness of SCATTER does not depend on a specific embedding model. These results indicate that the observed improvements are robust and not artifacts of evaluator–trainer embedding alignment.
\begin{table}[h]
    \centering
    \resizebox{0.85\linewidth}{!}{%
    \begin{tabular}{lc}
        \toprule
        \textbf{Hyperparameter} & \textbf{Value} \\
        \midrule
        Training Batch Size & 480 \\
        PPO Mini-Batch Size & 480 \\
        PPO Micro-Batch Size (per GPU) & 32 \\
        Total Epochs & 6 \\
        Group Rollout Size ($G$) & 5 \\
        Learning Rate & $3 \times 10^{-5}$ \\
        KL Loss & True \\
        KL Coefficient ($\beta$) & 0.001 \\
        LoRA & True \\
        LoRA Rank ($r$) & 64 \\
        LoRA Alpha ($\alpha$) & 32 \\
        Max Prompt Length & 1024 \\
        Max Response Length & 512 \\
        Generation Temperature & 1.0 \\
        \bottomrule
    \end{tabular}}
    \caption{Training configurations for GRPO and SCATTER.}
    \label{tab:config}
\end{table}
\subsection{Impact of similarity thresholds}
We further perform a comprehensive sensitivity analysis over multiple similarity thresholds. Specifically, we vary the matching threshold $\tau_{sr}, \tau_{sp} \in \{0.70, 0.80, 0.90\}$  for SP@16 and SR@16, and $\tau_{valid} \in \{0.40, 0.50, 0.60\}$ for VR@16. The detailed results are reported in Table~\ref{tab:sp_sr_threshold} and Table~\ref{tab:vr_threshold}. Across all threshold configurations, SCATTER consistently outperforms the GRPO baseline. These results demonstrate that the observed improvements are stable and not dependent on a specific similarity cutoff.
\begin{table*}[h]
\centering
\small
\setlength{\tabcolsep}{4pt}
\begin{tabular}{lccc|ccc}
\toprule
\multicolumn{1}{c}{\multirow{2}{*}{\textbf{Method}}} 
& \multicolumn{3}{c}{\textbf{OpenForecast}} 
& \multicolumn{3}{c}{\textbf{OpenEP}} \\ \cmidrule{2-7}
\multicolumn{1}{c}{} 
& \textbf{SP@1/SP@16} 
& \textbf{SR@1/SR@16} 
& \textbf{VR@16}  
& \textbf{SP@1/SP@16} 
& \textbf{SR@1/SR@16} 
& \textbf{VR@16} \\ 
\midrule

\textbf{Base} 
& $9.33^{\textcolor{darkgreen}{\pm 1.14}}/23.90$ 
& $4.69^{\textcolor{darkgreen}{\pm 0.66}}/13.04$
& $4.44$
& $7.01^{\textcolor{darkgreen}{\pm 1.74}}/18.89$
& $3.34^{\textcolor{darkgreen}{\pm 1.11}}/9.62$
& $4.09$ \\

\textbf{+ SFT} 
& $10.35^{\textcolor{darkgreen}{\pm 0.86}}/28.95$
& $5.34^{\textcolor{darkgreen}{\pm 0.56}}/16.31$
& $8.57$
& $4.44^{\textcolor{darkgreen}{\pm 2.19}}/23.33$ 
& $1.36^{\textcolor{darkgreen}{\pm 0.69}}/9.83$
& $6.82$ \\

\textbf{+ GRPO} 
& $18.16^{\textcolor{darkgreen}{\pm 0.55}}/24.56$
& $9.41^{\textcolor{darkgreen}{\pm 0.32}}/13.37$
& $0.64$
& $14.10^{\textcolor{darkgreen}{\pm 0.94}}/17.78$ 
& $4.96^{\textcolor{darkgreen}{\pm 0.53}}/6.42$ 
& $0.62$  \\

\textbf{+ SCATTER} 
& $17.65^{\textcolor{darkgreen}{\pm 1.32}}/43.64$ 
& $9.46^{\textcolor{darkgreen}{\pm 0.70}}/25.48$
& $10.95$
& $7.36^{\textcolor{darkgreen}{\pm 2.18}}/27.78$
& $2.36^{\textcolor{darkgreen}{\pm 0.94}}/10.99$ 
& $8.88$ \\  

\bottomrule
\end{tabular}
\vspace{-3pt}
\caption{Performance comparison on the Qwen3-7B-Instruct backbone. The green superscript indicates the standard deviation.}
\vspace{-6pt}
\label{tab:qwen3_7b_results}
\end{table*}
\subsection{Training Configuration}
\label{app:configuration}
We benchmark our method against four baselines: gpt-4o-mini, the non-finetuned base model, standard SFT, and standard GRPO. To ensure a fair evaluation, we enhance the prompt of base model and gpt-4o-mini with specific hypothesis constraints (see Appendix~\ref{app:prompt}). For the SFT baseline, we curate a dataset of 3k instances via knowledge distillation from GPT-4o-mini; to guarantee high-quality supervision, we employ a rewrite-and-replace strategy where the ground truth is stylistically adapted to replace the most semantically similar generated candidate. Notably, both the GRPO baseline and our method adopt a two-stage training paradigm: an initial SFT warm-up on these 3k instances to enforce output formatting, followed by the respective RL fine-tuning stages on 10K randomly sampled instances. The detailed configuration is shown in Table~\ref{tab:config}. During inference, we employ a rollout temperature of 0.7, nucleus sampling with $p=0.8$, top-$k$ sampling with $k=20$, and a maximum response length of 512 tokens.

\subsection{Impact of Model Size}
We further conduct experiments on Qwen2.5-7B-Instruct. As shown in Table~\ref{tab:qwen3_7b_results}, despite the stronger base capability of the larger backbone, SCATTER continues to provide consistent improvements over the corresponding baselines, indicating that the proposed reward design remains effective as model capacity increases. We acknowledge that evaluation on 70B-scale models would be valuable future work; however, due to computational and API constraints, such experiments are currently infeasible. Nevertheless, the consistent gains observed across both 3B and 7B backbones suggest that SCATTER’s reward formulation is largely backbone-agnostic and benefits from increased model capacity. Notably, the 7B model performs slightly worse than the 3B model on OpenEP. This may be due to the relatively small size of OpenEP, which can increase variance for larger models, as well as potential cross-domain overfitting since training is conducted on OpenForecast while evaluation is performed on OpenEP. These observations are preliminary, and a more systematic cross-scale and cross-domain analysis is left for future work.

\subsection{Impact of Sampling Rounds on Llama}
\label{app:sampling_llama}
\begin{figure}[t]
  \includegraphics[width=\columnwidth]{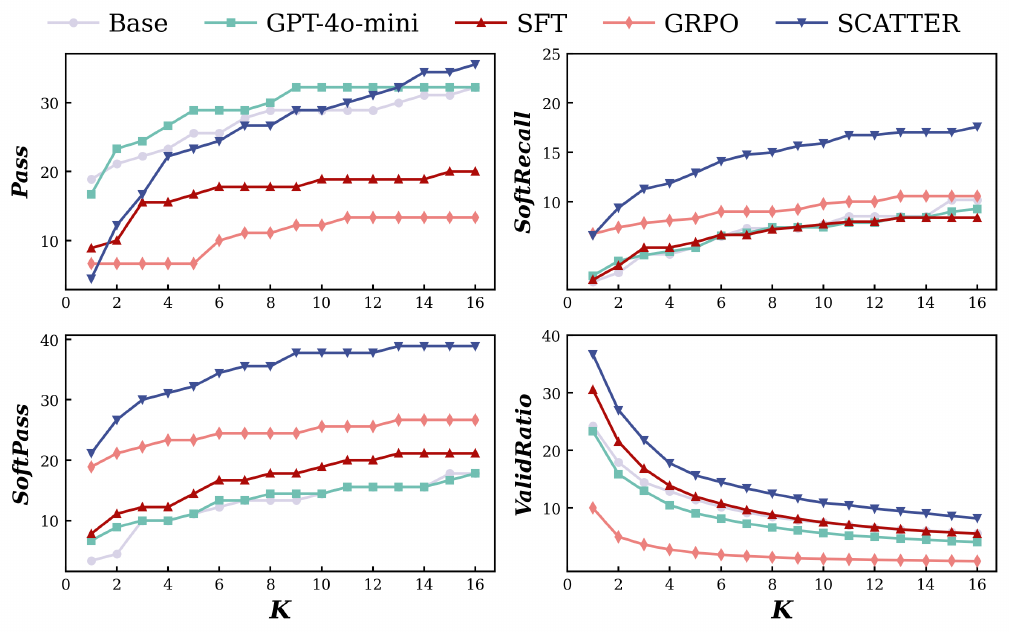}
  \caption{\textbf{Impact of sampling budget $K$ on generation performance.} Comparison of SCATTER against baselines on OpenForecast using llama3.2-3B-Instruct as Base Model. }
  \label{fig:sampling_openep}
\end{figure}

The  performance scaling of llama3.2-3B-Instruct across varying sampling rounds on OpenForecast are shown in Figure~\ref{fig:sampling_openep}.

\subsection{Case Study}
\label{app:case_study}
We proceed to illustrate an example from OpenEP, detailing the question, background, and generated hypotheses from a randomly sampled round as follows:
\begin{BlackPromptBox}{Question \& Background}
\textbf{Question}:\\
2024-07-05: What key developments can be expected in the dialogue between Viktor Orbán and Vladimir Putin regarding Ukraine's situation?
\\[1em]
\textbf{Background}:\\
Hungarian Prime Minister Viktor Orbán recently visited Kyiv for the first time since the onset of Russia’s full-scale invasion in 2022, sparking significant discussions with Ukrainian President Volodymyr Zelensky. During his visit, Orbán proposed prioritizing a ceasefire to expedite peace negotiations, a suggestion that contrasts with the ongoing European support strategy for Ukraine, which leans towards military aid. This proposal was set against the backdrop of Orbán's contentious relationship with EU policies due to his close ties with Russian President Vladimir Putin, whom he is slated to meet soon after his trip to Kyiv. 

While in Ukraine, Orbán emphasized the urgency of addressing the war, which he termed as Europe's "most important issue," suggesting that a ceasefire could accelerate peace talks. However, Zelensky maintained a firm stance, appreciating Hungary's humanitarian efforts but dismissing the ceasefire approach, which aligns with his previous rejections of similar proposals from other European leaders. Orbán's visit coincides with Hungary's recent assumption of the EU Council's rotating presidency, adding layers of diplomatic complexity given his government's contentious EU relations and internal democratic criticisms.

Orbán's outreach to Ukraine also included discussions on improving bilateral relations and economic cooperation, despite his history of blocking EU efforts to extend more robust support to Kyiv. This diplomatic maneuver occurs as Orbán gears up for strategic dialogues in upcoming high-profile European meetings, where Ukraine's geopolitical status and EU relations will be key topics. The Hungarian leader's actions and statements during this critical visit highlight the delicate balance of regional alliances, the contentious dynamics within the EU, and the ongoing challenges in the peace process amidst Europe’s most severe security crisis in decades.\\

\textbf{Ground Truth}:

[Viktor Orbán met Russian President Vladimir Putin.,

Orbán said Europe’s “war policy” is hurting its people and economy.,

The meeting with Putin was an unsanctioned attempt to resolve the conflict in Ukraine.]

\end{BlackPromptBox}

\begin{BlackPromptBox}{Hypotheses \hfill \textnormal{GPT-4o-mini}}
["Orbán may advocate for a stronger Russian influence in Ukraine during discussions with Putin.",\\
"Putin could propose increased military support for Hungary as a counter to EU sanctions.",\\
"Orbán might suggest a new economic partnership with Russia to bypass EU restrictions.", \\
"Putin could express discontent with Orbán's ceasefire proposal, urging a more aggressive stance.", \\
"Orbán may negotiate for energy supplies from Russia to stabilize Hungary's economy amid EU tensions.", \\
"Putin might offer concessions to Hungary in exchange for Orbán's support in the EU.", \\
"Orbán could push for a united front with Russia against Western sanctions.", \\
"Putin may highlight Hungary's importance as a strategic ally in Central Europe.", \\
"Orbán could seek assurances from Putin regarding Hungary's security amidst EU pressures.", \\
"Putin might leverage Orbán's EU presidency to influence European policies favorably towards Russia."] \\
\end{BlackPromptBox}

\begin{BlackPromptBox}{Hypotheses \hfill \textnormal{Qwen2.5-3B-Instruct}}
[Orbán may propose a joint statement condemning NATO expansion., \\
Putin could insist on troop withdrawal from eastern Ukraine territories., \\
Orbán suggests increased economic sanctions against Russia as leverage., \\
Putin might counter with demands for Ukrainian territorial integrity., \\
Orbán advocates for a neutral buffer zone along the Ukrainian-Russian border., \\
Putin proposes direct military assistance to Kyiv in exchange for political concessions., \\
Orbán calls for a temporary truce to focus on humanitarian aid and reconstruction., \\
Putin insists on a full-scale ceasefire to stabilize the region immediately., \\
Orbán seeks to restart stalled peace talks with renewed commitment., \\
Putin demands immediate recognition of separatist regions in Ukraine.]
\end{BlackPromptBox}
\begin{BlackPromptBox}{Hypotheses \hfill \textnormal{SCATTER}}
<answer>

["Ukrainian President Volodymyr Zelensky and Hungarian Prime Minister Viktor Orbán engage in a second summit in Berlin, Germany., \\
Orbán proposes a renewed push for a ceasefire in Ukraine amid ongoing tensions., \\
Zelensky maintains a firm stance against any ceasefire agreements, citing EU support.,\\
The Hungarian government assumes leadership of the EU Council's presidency starting July 5, 2024., \\
Orbán emphasizes the urgency of addressing the war, positioning it as Europe's most critical issue., \\
A news conference occurs in Budapest, Hungary, following Orbán's return from Kyiv., \\
Orbán and Zelensky agree on the need for improved bilateral relations and economic cooperation., 
The EU Council votes to extend sanctions against Russia, potentially complicating future negotiations., \\
Orbán's visit leads to a shift in public opinion within Hungary regarding the war and EU support., A third summit takes place in Brussels, Belgium, featuring Orbán and Zelensky alongside other European leaders.]

</answer>
\end{BlackPromptBox}

\subsection{Prompts}
\label{app:prompt}
\begin{PromptBox}{Prompt Templates for Post-Training}
\small 
You are an expert in event forecasting.
Given a forecasting question (Forecasting Question) and its background information (Background), your task is to generate 10 possible hypotheses.
\\[1em]
Here is the Background:\\
\{background\}

Here is the Forecasting Question:\\
\{question\}
\end{PromptBox}

\begin{PromptBox}{Prompt Templates for Evaluation}
\small 
You are an expert in event forecasting. Your task is to evaluate whether the Model Predictions correctly answers the Question, based on the Ground Truth (Correct Answer) provided.
The answer should be true if at least one item in Ground Truth is contained within the Model Predictions, otherwise false.
\\[1em]
Here is the Question:\\
\{query\}

Here is the Model Predictions:\\
\{model\_output\}

Here is the Ground Truth:\\
\{ground\_truth\}
\end{PromptBox}

\begin{PromptBox}{Prompt Templates for the Backbone Model}
\small 
You are an expert in event forecasting.
Given a forecasting question (Forecasting Question) and its background information (Background), your task is to generate possible hypotheses.
\\[1em]
Guidelines:

- Validity: Each hypothesis must be directly relevant to the Forecasting Question.

- Coverage: The set of hypotheses must collectively span different possible outcomes. Each hypothesis should be distinct from the others, avoiding redundancy or near-paraphrases edge cases.
\\[1em]
Here is the Background:\\
\{background\}

Here is the Forecasting Question:\\
\{question\}
\\[1em]
Constraints:
    
- Output exactly 10 new hypothesis.

- Each hypothesis should be specific and concise ($\le$ 25 words).

- Do NOT reference or rely on any information dated after \{date\}.

- Output must follow the JSON schema below and contain no extra text.
\\[1em]
Example output format:

["Prediction 1", "Prediction 2", "Prediction 3","Prediction 4",...,"Prediction 9","Prediction 10"]
\end{PromptBox}

\begin{PromptBox}{Prompt Templates for Answer Generation}
\small 
You are an expert in event forecasting. You are given a list of Ground Truth events (Ground Truth) and a list of model-generated hypotheses (Hypotheses). Your task is to rewrite each Ground Truth into a new hypothesis, following the same style, length, and tone as the Hypotheses. \\
\\[1em]
Guidelines:\\
- Preserve the factual meaning of each Ground Truth, but phrase it in the same stylistic form as the Hypotheses\\
- Hypotheses are typically concise, forward-looking statements ($\le$ 25 words).\\
- Maintain neutrality and avoid adding explanations, reasoning, or extra context.\\
- Ensure the wording is similar in style to the provided Hypotheses (e.g., same verb tense, tone, and abstraction level).\\
- Each Ground Truth must correspond to exactly one new hypothesis.\\
\\[1em]
Here is the Hypotheses:\\
\{hypotheses\}

Here is the Ground Truth:\\
\{ground\_truth\}\\
\\[1em]
Constraints:\\
- Do NOT invent new events beyond the Ground Truth.\\
- Output must be a JSON array of hypotheses, with one entry per Ground Truth, in the same order as the Ground Truth list.\\
\\[1em]
Output JSON schema:

[the number of Ground Truth here]
\end{PromptBox}

\begin{PromptBox}{Prompt Templates for Answer Extraction}
\small 
You are an expert in event forecasting. Your task is to convert a raw, unstructured answer paragraph into a structured list of concise "ground truth" statements. Given a Background, Question and Answer, extract the core factual points solely from the "Answer" text that directly answer the question.\\
\\[1em]
Guidelines:\\
- Remove conversational filler (e.g., "Indeed," "On the other hand," "According to a survey").\\
- Each statement should be concise (aim for $\le$25 words) but preserve key entities and numbers.\\
- Statements must be derived strictly from the "Answer" section. Do NOT include or infer information found only in the "Background", even if it provides context. Do NOT invent new events.\\
- Each Ground Truth must directly relate to answering the specific Question provided.\\

Here is the Background:\\
\{background\}

Here is the Question:\\
\{question\}

Here is the Answer:\\
\{answer\}
\\[1em]
Output JSON schema:

["Ground Truth 1","Ground Truth 2",...]
\end{PromptBox}

\end{document}